\documentclass[aps,pra,twocolumn,amsmath,amssymb,nofootinbib,showpacs,superscriptaddress]{revtex4-1}
\usepackage[english]{babel}
\usepackage{latexsym}
\usepackage{graphics}
\usepackage{graphicx}
\usepackage{epsfig}
\usepackage{color}
\usepackage{bm}
\usepackage{amsmath}
\usepackage{amssymb}
\usepackage{amsthm}
\usepackage{dcolumn}
\usepackage{bm}
\usepackage{float}
\usepackage{hyperref}
\usepackage{color}
\usepackage{epstopdf}
\usepackage{cleveref}
\usepackage[svgnames]{xcolor}
\hypersetup{hidelinks,colorlinks=true,allcolors=DarkBlue}

\begin{document}

\preprint{APS/123-QED}

\title{Modular quantum key distribution setup for research and development applications}

\author{V.E. Rodimin}
\affiliation{Russian Quantum Center, Skolkovo, Moscow 143025, Russia}
\affiliation{QRate, Skolkovo, Moscow 143025, Russia}
\author{E.O. Kiktenko} 
\affiliation{Russian Quantum Center, Skolkovo, Moscow 143025, Russia}
\affiliation{Steklov Mathematical Institute of Russian Academy of Sciences, Moscow 119991, Russia}
\author{V.V. Usova}
\affiliation{Russian Quantum Center, Skolkovo, Moscow 143025, Russia}
\affiliation{Skolkovo Institute of Science and Technology, Moscow 121205, Russia}
\author{M.Yu. Ponomarev}
\affiliation{Russian Quantum Center, Skolkovo, Moscow 143025, Russia}
\affiliation{QRate, Skolkovo, Moscow 143025, Russia}
\author{T.V. Kazieva}
\affiliation{Russian Quantum Center, Skolkovo, Moscow 143025, Russia}
\affiliation{QRate, Skolkovo, Moscow 143025, Russia}
\author{A.V. Miller}
\affiliation{Russian Quantum Center, Skolkovo, Moscow 143025, Russia}
\affiliation{QRate, Skolkovo, Moscow 143025, Russia}
\author{A.S. Sokolov}
\affiliation{Russian Quantum Center, Skolkovo, Moscow 143025, Russia}
\author{A.A. Kanapin}
\affiliation{Russian Quantum Center, Skolkovo, Moscow 143025, Russia}
\author{A.V. Losev}
\affiliation{Russian Quantum Center, Skolkovo, Moscow 143025, Russia}
\affiliation{QRate, Skolkovo, Moscow 143025, Russia}
\author{A.S. Trushechkin}
\affiliation{Russian Quantum Center, Skolkovo, Moscow 143025, Russia}
\affiliation{Steklov Mathematical Institute of Russian Academy of Sciences, Moscow 119991, Russia} 
\author{M.N. Anufriev}
\affiliation{Russian Quantum Center, Skolkovo, Moscow 143025, Russia}
\author{N.O. Pozhar}
\affiliation{Russian Quantum Center, Skolkovo, Moscow 143025, Russia}
\author{V.L. Kurochkin}
\affiliation{Russian Quantum Center, Skolkovo, Moscow 143025, Russia}
\affiliation{QRate, Skolkovo, Moscow 143025, Russia}
\author{Y.V. Kurochkin}
\affiliation{Russian Quantum Center, Skolkovo, Moscow 143025, Russia}
\affiliation{QRate, Skolkovo, Moscow 143025, Russia}
\author{A.K. Fedorov}
\affiliation{Russian Quantum Center, Skolkovo, Moscow 143025, Russia}
\affiliation{QRate, Skolkovo, Moscow 143025, Russia}

\date{\today}
\begin{abstract}
Quantum key distribution (QKD), ensuring the unconditional security of information, attracts a significant deal of interest.  
An important task is to design QKD systems as a platform for education as well as for research and development applications and fast prototyping new QKD protocols.
Here we present a modular QKD setup driven by National Instruments (NI) cards with open source LabView code, 
open source Python code for post-processing procedures,
and open source protocol for external applications.
An important feature of the developed apparatus is its flexibility offering possibilities to modify optical schemes as well as prototype and verify novel QKD protocols.
Another distinctive feature of the developed setup is the implementation of the decoy-state protocol, which is a standard tool for secure long-distance quantum communications.
By testing the plug-and-play scheme realizing BB84 and decoy-state BB84 QKD protocols, 
we show that developed QKD setup shows a high degree of robustness beyond laboratory conditions. 
We demonstrate the results of the use of the developed modular setup for QKD experiments in the urban environment. 
\end{abstract}

\maketitle
\newpage

\section{Introduction} 

In the view of a possible appearance of universal quantum computers in the next decades, 
the only way to ensure long-term information security is to use tools providing unconditional security such as the one-time pad scheme~\cite{Vernam,Shannon,Schneier}.~The 
crucial obstacle towards the implementations of unconditional security mechanisms is the key distribution problem~\cite{Schneier}. 

An elegant way, which allows overcoming this difficulty, is to use QKD~\cite{BB84,Gisin,Scarani,Lo}.
QKD provides a useful technique for the generation of private random keys between legitimate remote users (Alice and Bob).
It should be noted that in the classical world the mechanism preventing an eavesdropper (Eve) copying keys during their transmission is in fact absent. 
The crucial feature of the QKD technology is using of quantum objects as information carriers~\cite{BB84,Gisin,Scarani,Lo}. 
Due to the quantum no-cloning theorem~\cite{Wootters}, Eve is unable to keep a copy of sent quantum signals. 
Thus, QKD is a method offering information-theoretic security based on the physics laws. 
Remarkably, QKD is one of the first quantum technologies operating with individual quantum objects available as a commodity~\cite{Qiu}. 

However, an important demand is to design QKD platforms for research and development (R\&D) applications. 
They are meant for testing novel QKD protocols and demonstrations of attacks, as well as for educational purposes. 
A progressive platform for R\&D in quantum cryptography is Clavis~\cite{IDQ}, 
which is the generation of QKD setups by ID Quantique~\cite{IDQ}. 
This setup has been widely used in experiments on realization of novel QKD protocols, 
quantum hacking~\cite{Makarov2010,Makarov2015,Makarov2016}, 
testing post-processing algorithms~\cite{Lydersen2014}, 
quantum coin flipping~\cite{Lydersen2014},
and many others tasks.
Some of QKD setups have been designed, e.g. in order to show a cost-effective approach to the developing QKD systems~\cite{Larotonda2016,Rarity2006,Lo2014}.
Nevertheless, the decoy-state method~\cite{Hwang2003}, 
which can guarantee the security of the BB84 protocol~\cite{Hwang2003,LoMa2005,Wang2005,MaQiLo2005} and has already been realized in most of QKD setups, 
is still missing in QKD systems for R\&D and educational purposes. 
This lack limits the capabilities of such systems significantly.

Here we present a modular QKD setup for education, research and development applications. 
The crucial features of the developed apparatus are its modularity and flexibility offering possibilities to modify optical schemes,
and this versatility is a key difference from other research-oriented QKD setups available on the market. 
The versatility is essential for table-top realizations of new QKD technologies, 
which demand additional tools for quantum states manipulation~\cite{Weid} or hybrid quantum-classical information transmitting in a single fiber~\cite{Shields}. 
A distinctive feature of the developed setup is the implementation of the basic version of the decoy-state protocol~\cite{MaQiLo2005}.

The developed setup is driven by NI cards with open source LabView code~\cite{LabView, LabView2} for control and operating, open source Python code for the post processing~\cite{Kiktenko2016-Soft,Kiktenko2016-Soft2},
and open source protocol for external applications~\cite{Pozhar2018}. 
This post-processing procedure has been used for the key distillation in urban QKD experiments~\cite{Duplinskiy2018}.
The QKD setup can operate with any type of single-photon detectors (for details, see Refs.~\cite{IDQ,Kurochkin2017,Kurochkin2014,Kurochkin2015}) including superconducting single-photon detectors~\cite{Kurochkin2015}. 
The external drivers of single-photon detectors, phase modulators, and synchronization detectors are realized as removable modules. 
Each device can drive up to 8 detectors, 7 slow ports (VOA and etc.) and 5 universal ports for lasers, phase or amplitude modulators. 
The software solution in charge of controlling the system is written with the use of the development environment LabVIEW~\cite{LabView}. 
It is a popular tool used by scientists and others, mostly for data acquisition, instrument control, and industrial automation~\cite{LabView2}. 
We then consider LabVIEW as an optimal choice for educational purposes as well as for R\&D. 

Our paper is organized as follows.
We start by describing the developed QKD setup in Sec.~\ref{sec:modular}.
In Sec.~\ref{sec:plug}, we present the results of testing of the simplified version of the plug-and-play QKD scheme for laboratory and urban conditions.
In Sec.~\ref{sec:decoy}, we demonstrate operating of the developed QKD with the decoy-state QKD protocol. 
In Sec.~\ref{sec:post}, we describe the basics of the post-processing procedure with a focus on the analysis of the decoy-state statistics.
We summarize the main results of the present work and discuss further prospects in Sec.~\ref{sec:conclusion}.

\section{Modular QKD setup}\label{sec:modular}

\begin{figure}[h!]
\begin{center}
\includegraphics[width=0.625\columnwidth]{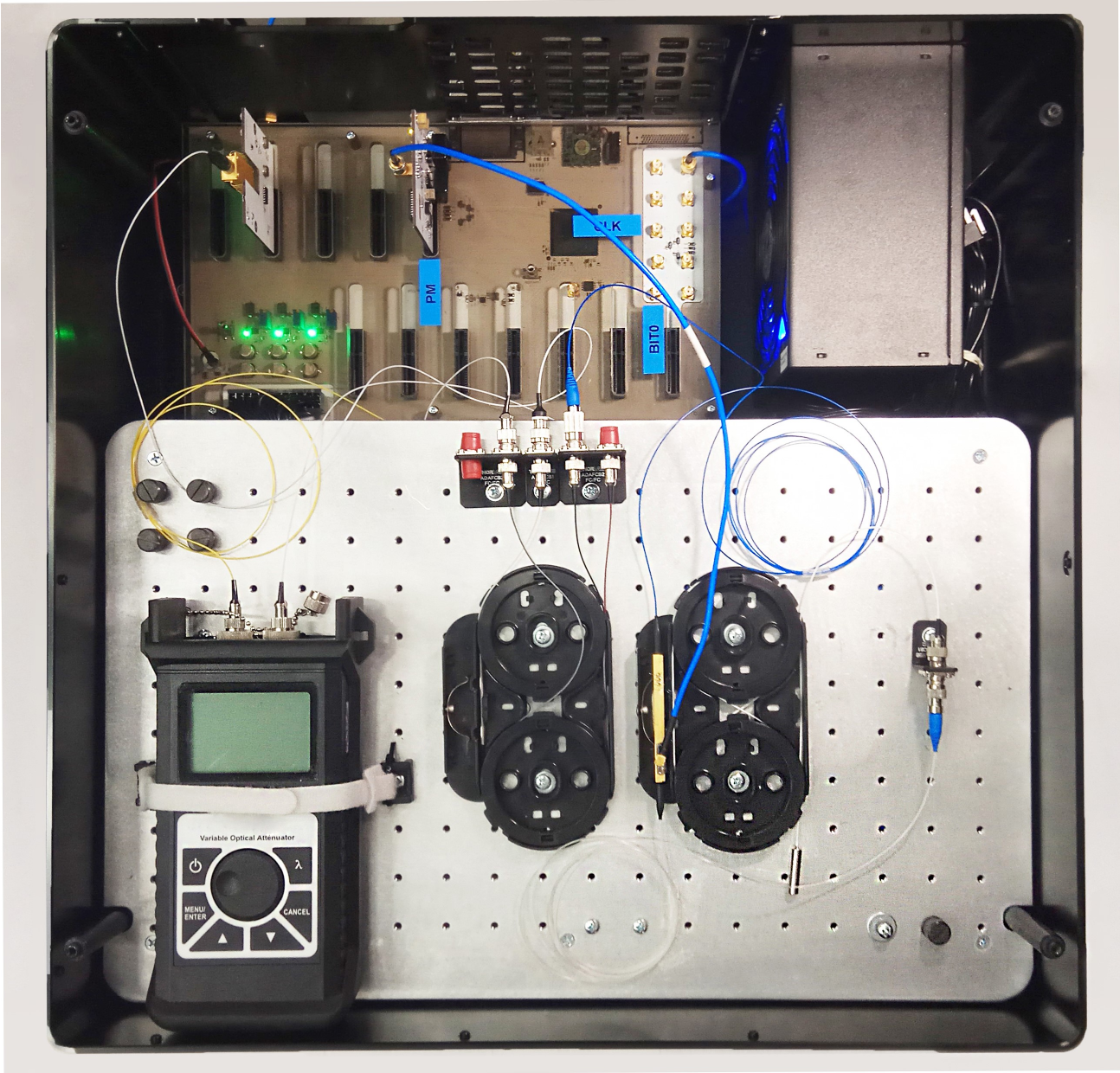}
\vskip -2mm
\caption{The developed modular QKD setup for R\&D applications and educational purposes}.
\label{fig:setup}
\end{center}
\end{figure}

The developed modular QKD setup consists of two essential components, Alice and Bob.
Alice is in charge of the quantum signals' preparation, whereas Bob measures the results. 
Fig.~\ref{fig:setup} shows a photo of Alice's unit (Bob's unit looks similarly).
On the top of the picture, one can see a motherboard with 10 SMA connectors and two socket rows for add-on cards. 
The laser module, phase, and amplitude modulator drivers are implemented in the form of such cards. 
The main function of the motherboard is to provide commutation between add-on cards and the NI board, installed in a PC and connected to the motherboard via a PCI cable. 
In addition, the motherboard implements the power supply control and executes a delaying of electric pulses to apply them at the correct time to optical and amplitude modulators. 

On the top side of the Bob or Alice unit, there is an optic table, where fiber optic components can be settled. 
Thus, the developed modular QKD setup is easily reachable from all sides.
One can then easily reconfigure both the optical scheme and the set of electronic driver cards. 
A sufficient amount of additional SMA connectors allows connecting an oscilloscope in order to investigate the shape of electrical pulses and their position on the timescale. 
These units are supplied with removable covers for transportation.

The details of the developed QKD setup are as follows.
The semiconductor laser LDI-DFB2.5G generates optical pulses 2 ns wide on the standard telecommunication wavelength 1.55 $\mu$m, the frequency is up to 10 MHz. 
The circulator, beamsplitters, Faraday mirror, and phase modulators are standard components. 
The quantum channel and storage line are single mode optical fibers.
Overall control of the electro-optic components is realized by NI PCIe-7811R installed in the corresponding PC (Bob or Alice). 
In total each device can drive up to 8 detectors, 7 slow ports (VOA and etc.), 10 SMA ports, 3 VHDCI ports, and 5 universal ports for lasers, phase or amplitude modulators supporting both parallel and serial interfaces. 

The sifted keys from the hardware engine of the modular QKD setup are the input for post processing.
The post-processing procedures of sifted keys are an inherent part of both industrial and R\&D oriented QKD systems. 
This procedure consists of information reconciliation, parameter estimation, and privacy amplification. 
The designed modular QKD setup uses proof-of-principle realizations of the procedure described in Refs.~\cite{Kiktenko2016,Kiktenko2017,Kiktenko2018,Kiktenko2018-2}. 
We describe the post-processing procedure in more details below (see Sec.~\ref{sec:post}).

\begin{figure*}[t]
\begin{center}
\includegraphics[width=1\linewidth]{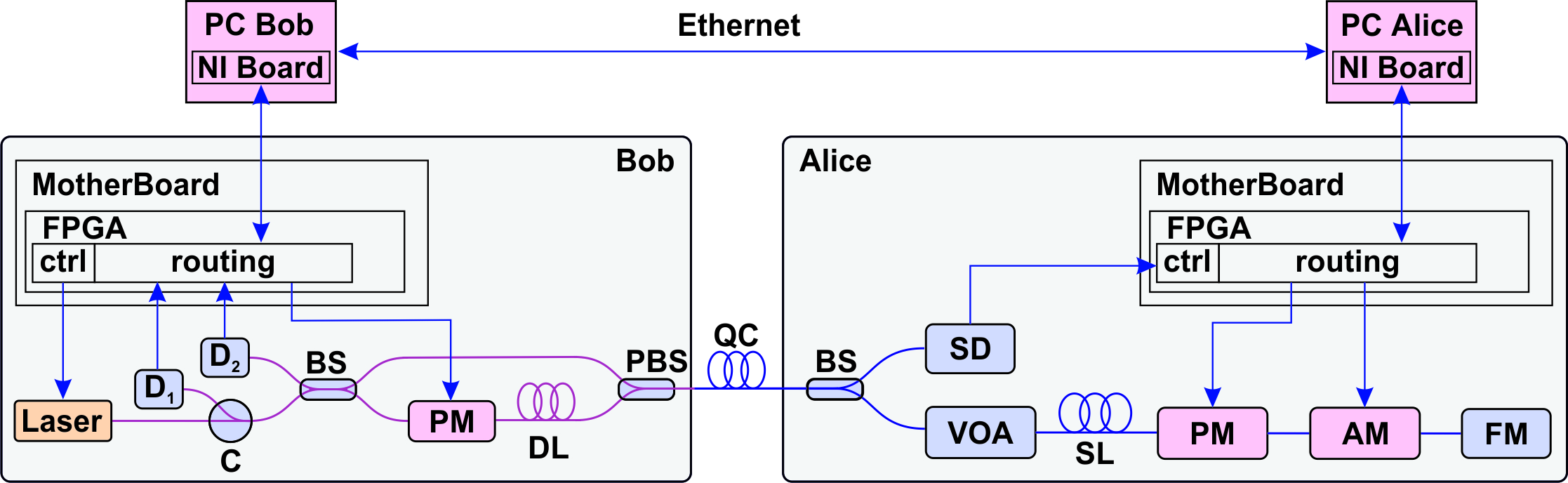}
\vskip -2mm
\caption
{The diagram of the developed setup. 
The used scheme is demonstrated: 
D1 and D2 are single-photon detectors, 
BS is the beamsplitter,
PBS is the polarization beamsplitter, 
PM is the phase modulator,
AM is the amplitude modulator,
VOA is the variable optical attenuator,
FM is the Faraday mirror, QC is the quantum channel,
SL is the storage line,
DL is the delay line,
SD is the synchro detector, 
C is the circulator, and PC is the personal computer. 
The operation is controlled by NI Boards connected via the Ethernet channel.}
\label{fig:setupQKD}
\end{center}
\end{figure*}

\section{Plug-and-play QKD scheme: BB84}\label{sec:plug}

We start tests of the developed setup from assembling a simplified version of the phase-encoding plug-and-play QKD scheme with the BB84 protocol without decoy states in various regimes (i.e. without the use of the amplitude modulators in our scheme, see Fig.~\ref{fig:setupQKD}). 
The used plug-and-play scheme, in this case, has a simple configuration. 
Thus, it is a perfect starting point for introductory studies in QKD (for details of the scheme, see Refs.~\cite{Gisin2,Kurochkin2011,Kurochkin2014,Kurochkin2015,Kurochkin2016}). 

\begin{table}[h]\center
\begin{tabular}{c|c|c}
\hline \hline Parameters &Laboratory conditions & Urban conditions \\
\hline $L$ & $25.5$ km&$30.6$ km \\
\hline ${\rm Attn}$ & $4.8$ dB&$11.7$ dB\\
\hline QBER & $1.6\% $&$5.1\% $  \\
\hline $R_{\rm rec}$ & $3.2$ kbit/s&$1.5$ kbit/s\\
\hline \hline
\end{tabular}
\caption
{\textbf{Implementation parameters and experimental results on realization of the plug-and-play scheme with the BB84 protocol}. 
Results of operating of the developed setup in laboratory and urban conditions,
where $L$ is the length of the communication channel and ${\rm Attn}$ is the total attenuation. 
The generation rate $R_{\rm rec}$ is for keys after information reconciliation.}
\label{table:plug}
\end{table}

The QKD procedure includes repetitive sessions of $10^3$ trains with $N_{\rm t}=2.4\times10^3$ pulses in each train. 
In our experiments with the plug-and-play QKD scheme. 
We use as short as possible time window in order to reduce the value of quantum bit error rate (QBER) in the laboratory conditions. 
The raw key goes on the sifting procedure after each session via the (authenticated) public channel by the TCP/IP protocol.
The data are obtained as a result of a five-day continuous operation in the laboratory, including unsupervised work. 
The output data are random 70.2 MB keys. 
We should point out that the operating of the setup is noisy, in particular, the standard deviation of QBER $\sigma({\rm QBER})=0.59\%$ is significant.

We also present the result of testing of the developed QKD setup with improved stability.
The results of the operating of the developed QKD setup are presented in Table~\ref{table:plug}.
After the stability improvement (both in hardware and software), the setup shows less noisy operation. 
During the 18 hours of intermittent work, 4 MB key was distributed with 1.6\% average QBER. 
The value of the standard deviation $\sigma({\rm QBER})=0.25\%$ in the stability regions is lower significantly than before ($\sigma({\rm QBER})=0.59\%$).

Although the fact that the developed QKD setup is oriented on education and research applications, 
it demonstrates a high degree of robustness beyond laboratory conditions. 
We present the results of the test of the developed setup between two bank offices in Moscow with the fully functional "plug-and-play" scheme.
The parameters of the experiment and the results of the operating of the developed QKD setup in the urban conditions are presented in Table~\ref{table:plug}.
The highest achievable sifted key generation rate is close to 2.2 kbit/s.
We should point out that this behavior is similar to the performance of IDQ's system~\cite{IDQ} and the recently presented QKD system~\cite{Larotonda2016}.
However, the fact that this data is obtained in urban conditions is important. 
In particular, in urban fibers lines, the effect of crosstalk on QBER can become significant~\cite{Kurochkin2016}. 
We also tested the setup for establishing a link in a heterogeneous quantum network in Moscow~\cite{Pozhar2017}. 

As it was mentioned before,
the important lack of such implementations of QKD setups for R\&D applications is missing decoy-states protocols.
That is the reason why for realistic experimental parameters we demonstrate the key rates after the information reconciliation stage, 
but not after privacy amplification (see Table~\ref{table:plug}).

\section{Decoy-state QKD protocol}\label{sec:decoy}

To exclude the shortcoming of our modular QKD setup regarding to missing the decoy-states method, we demonstrate the results of its implementation.
The decoy-state technique~\cite{Hwang2003,LoMa2005,Wang2005,MaQiLo2005} can guarantee the security of the BB84 protocol over long distances, 
and it has already been realized in most of QKD systems.
However, this method is not adopted in R\&D setups.

The decoy-state method is based on using laser pulses with different intensities. 
The intensities are chosen form a certain finite set. 
The choices for the pulses are kept in secret by Alice, 
but are publicly announced after the reception of all pulses by Bob. 
By analyzing (i) statistics of reception for pulses with different intensities and (ii) error rates for different intensities, 
one can estimate the fraction of single-photon pulses and the error rate for single-photon pulses. 
In particular, this allows detecting the photon number splitting attack~\cite{Hwang2003,LoMa2005,Wang2005,MaQiLo2005,CurtyLo2014,LimZbinden2014,Ma2017,Trushechkin2017}. 

\begin{table}[h]\center
	\begin{tabular}{c|c|c}
		\hline\hline
		Pulse type	 & Intensity & Probability of generation\\
		\hline
		Signal & $\mu=0.3$ & $p_\mu=0.5$	\\
		Decoy & $\nu=0.1$ & $p_\nu=0.25$	\\
		Vacuum & $\lambda=0.007$ & $p_\lambda=0.25$	\\
		\hline\hline
	\end{tabular}
	\caption{\textbf{Parameters of the decoy-state protocol.}} 
	\label{tab:decoy_param}
\end{table}

In our setup, we used a finite-key version of the decoy statistics analysis described in Ref.~\cite{Trushechkin2017}. 
We employ three types of pulses: ``signal pulse''  $\mu>0$ with the probability $p_\mu$,  ``weak decoy pulse'' $\nu>0$ with the probability $p_\nu$, and ``vacuum pulse'' $\lambda\geq0$ with the probability $p_\lambda=1-p_\mu-p_\nu$.
We note that the intensity of the ``vacuum state'' $\lambda$ is close to zero, but not exactly zero due to the technical reasons.
In the developed setup, it was at the level of $\lambda=0.007$.
In fact, the ``vacuum intensity'' is the second decoy intensity. 
It is required that $\lambda<\nu/2$ and $\lambda+\nu<\mu$. 
The parameters of the implemented decoy-state protocol are summarized in Table~\ref{tab:decoy_param}.

\begin{table}[h]\center
	\begin{tabular}{c|c}
		\hline\hline
		$L$ & 10 km \\
		Attn & 4.3 dB\\
		QBER & 2.1 \% \\\
		$R_{\rm sec}$ & 0.4 kbit/s \\
		\hline\hline
	\end{tabular}
	\caption{\textbf{Experimental results on realization of the decoy-state protocol.}
	The generation rates $R_{\rm sec}$ is for secure keys obtained after the privacy amplification stage given by Eq.~(\ref{eq:rate}).	
	} 
	\label{tab:decoy_rslts}
\end{table}

To test the performance of our decoy-state protocol we conducted an experiment in the laboratory conditions.
Again the QKD procedure includes repetitive sessions of $10^3$ trains with $N_{\rm t}=10^3$ pulses in each train.
The laser source under the control of the NI PCle-7841R board generates optical pulses 3 ns wide on the standard telecommunication wavelength 1.55 $\mu$m with the pulse repetition rate 5 MHz. 
Parameters of the experiment and resulting secure key rate are presented in Table~\ref{tab:decoy_rslts}.
Operating of the setup with the use of the BB84 QKD protocol with the use of decoy states is presented in Fig.~\ref{fig:QBER}.
The employed post-processing procedure, that is used for the secure key generation, is presented in the next section.

\begin{figure}[t]
\begin{center}
\includegraphics[width=1\columnwidth]{fig3.pdf}
\vskip -2mm
\caption
{Operating of the setup with the use of the BB84 QKD protocol with the use of decoy states. 
QBER is shown as the function of the bit block number (time). We demonstrate how the settings drift over time and after the adjusting procedure QBER returns to reasonable values.}
\label{fig:QBER}
\end{center}
\end{figure}

\section{Post-processing procedure}\label{sec:post}

Our post-processing procedure works as follows. 
Sifted keys go through the information reconciliation stage that is adjusted by the current value of QBER. 
This stage has two basic steps. 
The first is to use the LDPC syndrome coding/decoding~\cite{Gallager1962,MacKay1999} to correct discrepancies between keys using symmetric blind reconciliation~\cite{Kiktenko2016}. 
The universal polynomial hashing~\cite{Krovetz2001} aimed at the verification of the identity between keys after the previous step is used as the second one~\cite{Kiktenko2018}. 
As a result, Alice and Bob obtain a pair of verified keys which are identical up to verification error probability $\varepsilon_{\rm ec}$.
In our setup, we chose the parameters of employed polynomial hashing to have $\varepsilon_{\rm ec}=2\cdot10^{-11}$.

We note that the implemented information reconciliation procedure allows a significant increase in the efficiency of the procedure and reducing its interactivity~\cite{Kiktenko2016}. 
After the accumulation of a necessary number of blocks, the input goes to the parameter estimation. 
If an estimated value of QBER is higher than the critical value needed for efficient privacy amplification, the parties receive a warning message about possible eavesdropping. 
Otherwise, the verified blocks input privacy amplification stage, and estimated QBER is used in next rounds~\cite{Kiktenko2016,Kiktenko2016-Soft2}. 

The privacy amplification stage is used to reduce potential information of an adversary about the verified blocks to a negligible quantity. 
This is achieved by a contraction of the input bit string into a shorter string. 
The first problem of this stage is to determine the length $l_{\rm sec}$ of the final secret key (for each verified block) that provides the desired upper bound 
$\varepsilon_{\rm pa}$ of the failure probability of this stage. 
We have adopted the value $\varepsilon_{\rm pa}=10^{-12}$. 

Adaptation of formulas from Refs.~\cite{Tomamichel2017, Tomamichel2012} to our case leads to the following result:
\begin{equation}\label{eq:pa}
  l_{\rm sec}=\hat\kappa_1^{\rm l} l_{\rm ver}[1-h(\hat e_1^{\rm u})]-{\rm leak}_{\rm ec}+5\log_2\varepsilon_{\rm pa}.
\end{equation}
Here $l_{\rm ver}$ is the length of the verified block, ${\rm leak}_{\rm ec}$ is the number of bits of information about the verified block leaked to Eve during the information reconciliation stage, 
\begin{equation}
	h(x)=-x\log_2x-(1-x)\log_2(1-x)
\end{equation}
is the binary entropy function. 
We should point out that the main difference is that, QBER in our scheme is calculated after the error correction stage, not by random sampling from the sifted keys.

Further, $\hat\kappa_1^{\rm l}$ is a lower bound on the number of positions (bits) in the verified block obtained from single-photon pulses and $\hat e_1^{\rm u}$ is an upper bound the fraction of errors in such positions (corrected on the error correction stage),
The estimation of $\hat\kappa_1^{\rm l}$ and $\hat e_1^{\rm u}$ is given by the processing the decoy state statistics, 
which is described in Ref.~\cite{Trushechkin2017}. 
The quantity $h(\hat e_1^{\rm u})$ determines the potential knowledge of the bits obtained from single-photon pulses by the eavesdropper. 
This reflects the essence of QKD: it is impossible to get knowledge of the bits of the sifted key obtained from single-photon pulses without introducing errors in them. 

If $l_{\rm sec}$ given by Eq.~(\ref{eq:pa}) is positive, then the secret key distribution is possible for the considered verified block. 
If $\tau$ is the time needed to generate a verified block with the length $l_{\rm ver}$, then the secret key rate is as follows:
\begin{equation}\label{eq:rate}
	R_{\rm sec}=l_{\rm sec}/\tau.
\end{equation}

After the calculation of the length of a final key (for each verified block), the privacy amplification itself is performed: 
the block of the final key is computed as a hash function of the verified block.  
The family of hash functions is required to be 2-universal.
In the developed setup, we employed a Toeplitz hashing for this purpose.

\section{Discussion and outlook}\label{sec:conclusion}

We have present the modular QKD setup for R\&D applications and educational purposes. 
An important feature of the developed apparatus is its flexibility offering possibilities for modifying optical schemes and testing novel QKD protocols. 
Another distinctive feature of the developed setup is the implementation of the decoy-state protocol. 
We have demonstrated the results of testing the plug-and-play scheme realizing BB84 and decoy-state BB84 QKD protocols. 
We have also used the setup for distribution of keys between two bank offices in Moscow.

Due to the quantum basis and optical methods of information transfer, the developed modular QKD setup is well suited for the education of undergraduate students. 
Taking into account challenging tasks to generate, control, modify and synchronize electric and optical pulses, it is a powerful course for engineers. 
Involving the necessity to correct errors in sifted keys and privacy amplification, we get a wide field for research for mathematicians. 
Additional modules for preparation, manipulation, and measurements of quantum states allows one to investigate novel QKD protocols and attack scenarios. 

LabView code for control and operating is available on the public, free repository~\cite{LabView}.
Our source code for a proof-of-principle realization of the post-processing procedure is available~\cite{Kiktenko2016}.
We should also note that resulting keys can be used for a variety of external applications, and our protocol (client API) for possible applications is freely distributed as well~\cite{Pozhar2018}.

Further steps of our work include the study of parallelization of the developed post-processing procedure, investigation of a possible speed-up, 
and comparison with forefront QKD software projects such as R10 by AIT~\cite{AIT} with focus on creating QKD network solutions~\cite{SECOQC,Pozhar2017}.
In particular, we note that the presented scheme can be used a testing platform for novel tools for overcoming the detection-efficiency mismatch problem in the decoy-state protocol~\cite{Trushechkin2018,Ma2018}.
We also looking for an extensions of possible educational activities with this setup~\cite{FedorovRodimin2018}.

\section*{Acknowledgments} 

The early stage (2015-2017) of the present work on prototyping the device has been supported by Ministry of Education and Science of the Russian Federation in the framework of the Federal Program (Agreement 14.582.21.0009, ID RFMEFI58215X0009).
The advanced post-processing procedure and the implementation of the decoy-state protocol is supported by the Russian Science Foundation under Grant No. 17-71-20146 (work of V.E.R., E.O.K., V.L.K., Y.V.K., and A.K.F.).

\end{document}